# The role of cooling flows in galaxy formation

P.E.J. Nulsen[1] and A.C. Fabian,[2]
[1] Department of Physics, University of Wollongong, Wollongong NSW 2522, Australia
[2] Institute of Astronomy, Madingley Road, Cambridge CB3 0HA

**ABSTRACT**
The present structure of galaxies is governed by the radiative dissipation of the gravitational and supernova energy injected during formation. A crucial aspect of this process is whether the gas cools as fast as it falls into the gravitational potential well. If it does then rapid normal star formation is assumed to ensue. If not, and the gas can still cool by the present time, then the situation resembles that of a cooling flow, such as commonly found in clusters of galaxies. The cooled matter is assumed to accumulate as very cold clouds and/or low mass stars, i.e. as baryonic dark matter. In this paper we investigate the likelihood of a cooling flow phase during the hierarchical formation of galaxies. We concentrate on the behaviour of the gas, using a highly simplified treatment of the evolution of the dark matter potential within which the gas evolves. We assume that normal star formation is limited by how much gas the associated supernovae can unbind and allow the gas profile to flatten as a consequence of supernova energy injection. We find that cooling flows are an important phase in the formation of most galaxies with total (dark plus luminous) masses $\gtrsim 10^{12} \, M_\odot$, creating about 20 per cent of the total dark halo in a galaxy such as our own and a smaller but comparable fraction of an elliptical galaxy of similar mass. The onset of a cooling flow determines the upper mass limit for the formation of a visible spheroid from gas, setting a characteristic mass scale for normal galaxies. We argue that disk formation requires a cooling flow phase and that dissipation in the cooling flow phase is the most important factor in the survival of normal galaxies during subsequent hierarchical mergers.

**Key words:** clustering – cooling flows – galaxies: formation, elliptical and lenticular, spiral – galaxies: stellar content – X-rays: galaxies.

## 1 INTRODUCTION

Radiative cooling of gas is generally considered to be an important mechanism in the formation of the visible parts of galaxies. Following the work of White & Rees (1978), it is assumed that gas falls into potential wells of dark matter where it is heated and subsequently cools. The potential wells (dark haloes) evolve in a hierarchical manner such that present-day galaxies consist of several earlier and smaller objects which were in turn formed of smaller ones. The nature of the evolution of the dark matter depends on the spectrum of mass fluctuations from the early Universe (Peebles 1993).

Here we are interested primarily in the evolution of the gas rather than the dark haloes. A basic parameter for the gas is the ratio $\tau$ of the cooling time $t_{\rm cool}$ to the free-fall time $t_{\rm grav}$ (Rees & Ostriker 1977). If $\tau < 1$ then the gas is barely heated and it cools rapidly, collecting as clouds in which stars are assumed to form with a standard initial-mass function (IMF), as in the disk of the Galaxy. If $\tau > 1$, and $t_{\rm cool}$ is less than the age of the system, a cooling flow is formed. These are common in clusters of galaxies (see Fabian 1994 for a review) where cooling rates of tens to hundreds of solar masses per year are inferred from X-ray data. Observations at optical and other wavelengths indicate that the cooled gas there must remain as very small, very cold clouds and/or form low-mass stars.

The goal of this paper is to investigate how widespread a cooling flow phase might be during galaxy formation and what significance it could have for galaxy formation. A first attempt at this was carried out by Thomas & Fabian (1990), where is was assumed that a single value of $\tau$ could be assigned to a protogalaxy. Cooling flows were found to be an important phase in the formation of the most massive galaxies. In a later paper (Fabian & Nulsen 1994), we found that since $\tau$ increases with radius across any realistic protogalaxy, a halo cooling flow is likely even during the formation of our own Galaxy. Here we study the behaviour of the gas during the evolution of a highly simplified hierarchical clustering model, paying particular attention to making our treatment of the structure, metallicity and behaviour of the gas at each stage as realistic as possible.

An important ingredient of hierarchical models for galaxy formation is feedback due to heating from supernovae (Larson 1974; Dekel & Silk 1986). If this were not included much of the gas would cool and form stars in the first dark haloes and so be excluded from further evolution. The large baryon fraction in clusters of galaxies, of between 10 and 25 per cent (White & Frenk 1991; White et al 1993; White & Fabian 1995), which exceeds that in the stars of the member galaxies by a factor of about 2 to 5, testifies that this does not occur. We use a simplified model to follow the metal (iron) enrichment and heating of the gas by supernovae during the hierarchical growth of a galaxy. Heating distends the gaseous atmosphere, promoting a cooling flow phase, while metal enrichment enhances cooling, working against the formation of cooling flows.

Other workers have studied hierarchical models for galaxy formation with varying degrees of sophistication in dealing with the gas (White & Frenk 1991; Cole 1991; Navarro & Benz 1991; Lacey & Silk 1991; Kauffmann, White & Guiderdoni 1993; Cole et al 1994). These models all differ from ours in important details. In particular, none deals with the effect of the heat injected by supernovae on the gas in subsequent collapses. Instead, all of the semi-analytical models have assumed that collapse distributes the gas with the same isothermal profile as the dark matter. While numerical simulations which include supernova heating (*e.g.* Navarro & White 1993) include this effect, in principle, poor spatial resolution has limited their value (see section 2.2).

We follow our earlier work (Thomas & Fabian 1990; Fabian & Nulsen 1994) in assuming that the nature of the star formation changes when $\tau \gtrsim 1$ and a cooling flow operates (Kauffmann et al 1993 use the same assumption for the most massive objects in their models). Most significant, matter deposited by cooling flows is assumed to have a high mass-to-light ratio, *i.e.* it is baryonic dark matter.

In section 2 we describe the galaxy evolution model. In section 3 we give examples of the properties of the objects — galaxies and clusters of galaxies — which form in the model. In section 4 we discuss the results of our simulations, arguing that a cooling flow phase occurs in most "normal galaxy" formation, and that the occurrence of this phase has major consequences for the structure and evolution of galaxies.

## 2 A MODEL FOR GAS IN EVOLVING HALOES

### 2.1 The dark matter

The model we use for the evolving dark matter haloes assumes that they are isothermal spheres ($\rho(R) \propto R^{-2}$) formed by gravitational instability in a dust dominated Einstein–de Sitter Universe. The haloes are truncated at the radius where their mean density is 200 times the background density at the collapse epoch (see Cole 1991; Kauffmann et al 1993), *i.e.* for a protogalaxy of total mass $M$

$$R_0 = 0.1 H_0^{-1} (1+z)^{-3/2} v_c,$$

where $v_c$ is the Kepler circular velocity in the halo ($v_c^2 = GM/R_0$) and the Hubble constant is $H_0$ (taken as $50\,\mathrm{km\,s^{-1}\,Mpc^{-1}}$ in our calculations).

The haloes are assumed to grow in mass during a succession of $p$ hierarchical collapses at equal redshift intervals between $z_{\mathrm{start}}$ and $z_{\mathrm{fin}}$ (usually $z_{\mathrm{fin}} = 0$). Halo mass increases by the same factor at each successive collapse, generally starting from $10^{10}\,\mathrm{M}_\odot$ and ending at $10^{15}\,\mathrm{M}_\odot$ after $p$ collapses. In a realistic model for the haloes, the growth would follow a more probabilistic path depending on the spectrum of mass fluctuations in the Universe (see e.g. White & Frenk 1991). We choose here to leave the halo model simple and concentrate on the behaviour of the gas.

### 2.2 The gas

Since the mass fraction of baryons in clusters is high (10 to 25 per cent) we shall adopt a similar high fraction of baryons $f_B$ in our model. (This ignores constraints from calculations of cosmic nucleosynthesis in a flat Universe, eg Walker et al 1991. Otherwise there is no mechanism in our model which can account for the high baryon fraction in clusters.) The gas is initially assumed to collapse with the isothermal dark matter haloes and, in the absence of any non-gravitational heating, adopt an isothermal profile $\rho \propto R^{-2}$ out to $R_0$ at the virial temperature $T_{\mathrm{vir}} = GM\mu m_H/2kR_0$.

Gas for which $\tau < \tau_0$, where $\tau_0 \sim 1$, is assumed to cool instantly and begin forming stars with a normal IMF. We denote the radius at which $\tau = \tau_0$ as $R_{\mathrm{CF}}$. This is calculated assuming that $t_{\mathrm{cool}} = \frac{3}{2} n_T kT / n_e n_H \Lambda$ and $t_{\mathrm{grav}} = \frac{R}{v_c}\sqrt{\frac{\pi}{2}}$, where the total particle density in the gas is $n_T$, and the electron and proton densities are $n_e$ and $n_H$ respectively. The cooling function $\Lambda$ is a simple approximation to the metallicity dependent results given by Böhringer & Hensler (1989).

Using $\tau_0$ close to 1, our treatment is somewhat more favourable to the formation of cooling flows than that of White & Frenk (1991; also Kauffmann et al. 1993), who compared the cooling time of the gas to the collapse time of the halo. In the absence of radiative cooling, collapsing gas would be shocked and then heated to the virial temperature by (adiabatic) compression. Shock passage is generally fast, so that the outcome of the collapse is governed by competition between radiative cooling and compressive heating of the shocked gas. Since the collapsing gas does not generally move far after being shocked, the compression takes about one free-fall time at the place where the gas comes to rest, which is why we use $t_{\mathrm{grav}}$ in our criterion. We take the view that any gas which is heated close to the virial temperature before it starts to cool significantly takes part in a cooling flow. Note that Cole et al. (1994) only consider the total amount of gas that can cool, so that they do not distinguish catastrophic cooling from a cooling flow phase.

At each stage of the collapse hierarchy the infalling gas is likely to be inhomogeneous as a result of previous collapse stages. This means that in reality the hot and cold phases will interpenetrate near to $R_{\mathrm{CF}}$ (*i.e.* the gas forms 2 phases) immediately following the collapse. Because of the large temperature difference between the phases, the hot gas will occupy the great bulk of the volume. The separation into phases is favourable to forming a substantial cooling flow at the earliest opportunity. The relatively poor spatial resolution of numerical simulations (*e.g.* Evrard, Summers & Davis 1994; Kang et al. 1994) prevents them from modelling this situation well. In particular, poor resolution forces shocks to be made much thicker than they are in reality. Because radiative cooling is generally more effective at low temperatures, this greatly increases the opportunity for radiative cooling to counteract shock heating by allowing time for gas to cool during the heating phase of a shock.

Supernovae from the high mass stars, above 8 M$_\odot$, affect the remaining gas by enriching it with metals (followed here in terms of iron abundance), thereby increasing its radiative cooling rate, and by dumping energy into it (as heat and bulk kinetic energy). We assume that star formation proceeds in the gas from $R < R_{\rm CF}$ until either that gas is used up or the energy from the supernovae has just unbound the whole of the remaining atmosphere (Dekel & Silk 1986; Thomas & Fabian 1990). The fraction of the mass within $R < R_{\rm CF}$ needed to do this is denoted $f_{\rm unbind}$.

This differs from other recent treatments (*e.g.* White & Frenk 1991; Cole et al. 1994) where it is assumed that feedback from supernovae regulates the star formation rate rather than unbinding the gas. White & Frenk argue that the time between star formation and energy release by supernovae is much shorter than the timescale for star formation (comparable to the free-fall time, $t_{\rm grav}$). This is certainly true for supernovae from the most massive stars (although less massive stars may take as long as $3 \times 10^7$ y to produce a supernova). However, the energy injection required to prevent star formation is about equal to the energy required to raise the gas to the virial temperature (it cannot be much greater without ejecting the gas). As a result, the effects of the energy input propagate through the protogalaxy at about the sound speed of the hot gas, *i.e.* about the free-fall speed. Thus, supernova feedback over large scales occurs on roughly the same timescale as the star formation, not fast enough to tightly regulate the rate of star formation. In our view this situation is unstable, so that, where they can produce sufficient energy, the supernovae will drive the bulk of the gas out of a dark halo.

There is a further problem with the supernova-regulated star formation model used by White & Frenk (1991) (and their followers). They estimate the rate at which hot gas cools using the initial distribution of hot gas. However, when supernova heating prevents some of this gas from cooling, as required by their regulation mechanism, the gas accumulates. White & Frenk suggest that reheated gas is returned to a galactic halo via a fountain (although, to account for extra potential energy in addition to radiative losses, this requires somewhat more than the $v_c^2$ per unit mass which they allow). Unless this gas is ejected from the system, it continues to cool, and at some later stage supernova feedback must heat the accumulated gas together with the gas that would have been cooling in the absence of feedback. The heat required to prevent the accumulated gas from forming stars is not considered by White & Frenk (1991), who have therefore underestimated the total star formation in their models. This problem does not arise in models where the gas is ejected. It also casts doubt on their claim that supernova regulation leads to less star formation than in models where the atmosphere is ejected.

The mean mass of iron produced per supernova is 0.07 M$_\odot$ (see overview by Renzini et al 1993) which, with an energy release into the gas of $4 \times 10^{50}\epsilon_{\rm SN}$ per supernova ($\epsilon_{\rm SN} \sim 1$, Spitzer 1978), means a relationship between specific energy input and iron abundance $Z_{\rm Fe}$ of $5.7 \times 10^{15}\epsilon_{\rm SN} Z_{\rm Fe}/Z_\odot$ erg g$^{-1}$ ($Z_\odot = 0.002$ by mass for iron).

Supernova energy is deposited in the gas as heat or kinetic energy. However, gas which is ejected from a dark halo by the supernovae stores the additional energy principally as gravitational binding energy, in which form it can remain until a subsequent collapse. We assume that the supernova energy is retained in the gas until the next stage of collapse. The atmosphere formed at the next stage is isothermal but hotter than the virial temperature, causing it to have a shallower density profile $\rho \propto R^{-2\beta}$, where $\beta = \mu m_{\rm H} \sigma^2 / kT$ and the velocity dispersion $\sigma = v_c/\sqrt{2}$.

The first small objects which form have $M \sim 10^{10}$ M$_\odot$, $\beta = 1$ and $f_{\rm unbind} \ll 1$. In detail $f_{\rm unbind}$ is obtained by solving

$$f_{\rm unbind} \frac{M_{\rm gas}(R < R_{\rm CF})}{M_{\rm SN}} 4 \times 10^{50} \epsilon_{\rm SN} = \frac{5}{2}\sigma^2 \chi(\beta) M_{\rm gas}(R > R_{\rm CF})$$

$$+ (1 - f_{\rm unbind}) 4\sigma^2 M_{\rm gas}(R < R_{\rm CF}),$$

where $M_{\rm SN}$ is the mass of stars needed to form to make one supernova ($\sim 80$ M$_\odot$ for a standard IMF; Thomas & Fabian 1990), so $f_{\rm unbind} M_{\rm gas}(R < R_{\rm CF})/M_{\rm SN}$ is the total number of supernovae. The binding energy per unit mass of the hot gas is $5\chi\sigma^2/2$ and that of the catastrophically cooled gas is $4\sigma^2$ (ignoring self-gravity of the gas). Here $\chi(\beta)$ corrects the binding energy of the hot gas for $\beta \neq 1$ ($\chi(1) = 1$). The binding energy is calculated assuming that hot gas has the radial density profile and temperature given above, and is truncated at $R_0$.

Knowing $f_{\rm unbind}$ we can calculate the amounts of heavy metals and additional energy injected into the remaining gas. Gas which has cooled catastrophically has radiated away its thermal energy. This is taken into account when calculating the net excess energy in the gas.

The total energy per unit mass of the gas in an atmosphere with given $\beta$ exceeds that in an atmosphere with $\beta = 1$ by

$$\frac{E_{\rm excess}}{M} = \sigma^2 \frac{(9 + 2\beta)(1 - \beta)}{2\beta(3 - 2\beta)}.$$

To determine $\beta$ at the next stage of the hierarchy we equate this to the excess energy per unit mass in the gas determined as above. Solving for $\beta$ then enables us to calculate $R_{\rm CF}$ and $f_{\rm unbind}$ for the new atmosphere and, hence, the new iron and energy enrichment.

We thus step from small to large masses calculating the extent of the regions where gas cools catastrophically and of the cooling flow at each stage. We then estimate the fraction of the cooling flow gas which cools before the next collapse stage or cools by now (if the present collapse is the final point in the hierarchy). We also obtain the surviving mass fraction in gas and its metallicity at the end of the stage. For total masses of $10^{15}$ M$_\odot$ these can be compared with the baryon fraction and iron abundance of clusters today. For total masses of $10^{12}$ M$_\odot$, if the hierarchy stops at that mass, we can compare with the properties of our Galaxy and other "normal" galaxies.

As the hierarchy proceeds so $f_{\rm unbind}$ increases to unity, which means that the supernova energy cannot unbind the whole atmosphere in the more massive haloes. After this stage the supernovae due to star formation will heat the hot gas, affecting the amount of gas which cools from a cooling flow in the stage that they occur. We do not allow for this effect (but the supernova heat is carried through to the next stage). As discussed below, there is only one stage of the hierarchy where this is likely to make any appreciable difference.

For our basic results we assume that all the gas which cools from the cooling flow phase before the next collapse forms low-mass stars and that massive stars (and thus supernovae) only occur in gas which has cooled rapidly in the inner region ($R < R_{\rm CF}$).

In summary, the model details are calculated as follows:

(i) The collapse time and total mass of the system are computed from the collapse redshift. These determine the truncation radius, circular velocity, virial temperature, etc. Abundance and gas fraction are carried over from the end of the previous stage.

(ii) The excess energy is determined (from the abundance and radiative losses at previous steps) and this is used to determine $\beta$.

(iii) $R_{\rm CF}$ is calculated. This determines how much mass is available for normal star formation.

(iv) The fraction ($f_{\rm unbind}$) of the catastrophically cooled gas which must turn into stars in order to produce enough supernovae to unbind the remaining atmosphere is computed.

(v) $f_{\rm unbind}$ determines the mass of normal star formation and the enrichment of the remaining gas (for the next step).

(vi) If $f_{\rm unbind} = 1$ there is a cooling flow phase, and the amount of hot gas which cools before the next stage (or the present) is determined.

(vii) The fractions of matter left as gas, normal stars and baryonic dark matter at the next collapse can then be determined.

## 3 RESULTS

The results from our default model are shown in Fig. 1. Its parameters are $p = 9$ steps, starting baryon fraction $f_{\rm B} = 0.3$, $\tau_0 = 2$, $z_{\rm start} = 5$, $z_{\rm fin} = 0$, energy injection efficiency $\epsilon_{\rm SN} = 1$ and $M_{\rm SN} = 80\,{\rm M}_\odot$. A cooling flow phase develops rapidly for masses equal to and above $10^{12}\,{\rm M}_\odot$ and accounts in the end for more than 10 per cent of the total mass.

### 3.1 The Galaxy

Note that $10^{12}\,{\rm M}_\odot$ is approximately the total mass of our Galaxy (Kulessa & Lynden-Bell 1992; Norris & Hawkins 1991). If the hierarchy had stopped at $10^{12}\,{\rm M}_\odot$ objects, then all of the hot gas would have cooled into low-mass stars by now giving them baryonic dark haloes containing more than 20 per cent of the total mass. The spheroid, which we identify with the stars formed from the catastrophically cooled gas within $R_{\rm CF}$ (i.e. the normal stars with $M < 1\,{\rm M}_\odot$) contains about 8 per cent of the mass. The iron abundance is about 0.05 when the bulk of the spheroid stars form.

According to our model the spheroid stars form on a timescale comparable to the free-fall time, allowing very little time for the energy dissipation which is needed if the spheroid is to collapse any further. Thus, the size of the spheroid should be comparable to $R_{\rm CF}$. This also means that the early spheroids are not much more tightly bound than the dark haloes in which they are embedded, in which case they are prone to disruption at the next stage of the collapse hierarchy.

One problem for the model then, is that the bulk of the spheroid stars form at the step before the collapse which best represents the Galaxy. At that stage $R_{\rm CF}$ is close to 30 kpc, so that, although the total mass of the spheroid is about right, it would be much too large to represent that of the Galaxy. Indeed, a $10^{10}\,{\rm M}_\odot$ spheroid spread over a volume of 30 kpc radius would be quite difficult to detect.

How is it that $R_{\rm CF}$ can change so dramatically in one step of the hierarchy? At each step in the collapse hierarchy the temperature increases and the mean density decreases, both tending to increase cooling times. Along with this, the gas fraction decreases, driving the gas density down even faster. Finally, decreasing $\beta$ flattens the gas profile, making the ratio $\tau$ a less sensitive function of $R$ (in fact, for $\beta = 0.5$, $\tau$ is independent of $R$ and for smaller $\beta$ it is a decreasing function of $R$). Thus, for $\beta \simeq 0.6$, the location of $R_{\rm CF}$ is very sensitive to the other parameters. The net effect is that $R_{\rm CF}$ changes very rapidly as we proceed through the hierarchy.

We illustrate this sensitivity by running the same collapse history (i.e. all physical parameters the same and with the same total mass as a function of the time) but with the collapses taking place at different times (and hence masses). The result is illustrated in Fig. 2, where the collapses occur about midway between those of the default model. Progress of the collapse is generally very similar, apart from the formation of the spheroid. The collapse with mass $1.6 \times 10^{12}\,{\rm M}_\odot$ now produces a spheroid of about $5 \times 10^9\,{\rm M}_\odot$ inside about 5 kpc.

Although this sensitivity is partially an artifact of our treatment of the heat input, the factors affecting it are all real. This suggests that the mass of a galactic spheroid is sensitive to its history of formation.

We note that although the model clearly shows small spheroids forming even in the most massive systems, this is largely an artifact of the simple treatment we use. Once cooling flows form the (great bulk of) the gas involved in mergers will be hot, so that the gas in the merger product will probably have a flat core rather than a power-law density profile going to $R = 0$. A very small flat core will eliminate the region where the gas cools catastrophically and hence the spheroids in the late collapses.

As noted in the previous section, we have not allowed for heating of the hot gas by supernovae from the current step. In the early collapses there is no hot gas. In the late collapses the spheroid component is very small (non-existent in fact) and the gas temperature rising, so that the heat input can safely be ignored. The only stage of the collapse where supernova heating of the hot gas might be significant is at the first collapse to produce a cooling flow. This is the one occasion when the spheroid may be large enough to produce enough supernovae to cause substantial heating of the cooling flow gas.

### 3.2 Disks

Because of the low dissipation during spheroid formation there is little chance of disks forming from the pre-spheroid gas. We have also argued that the cooling flow phase produces low mass objects which are dark, so where do disks come from?

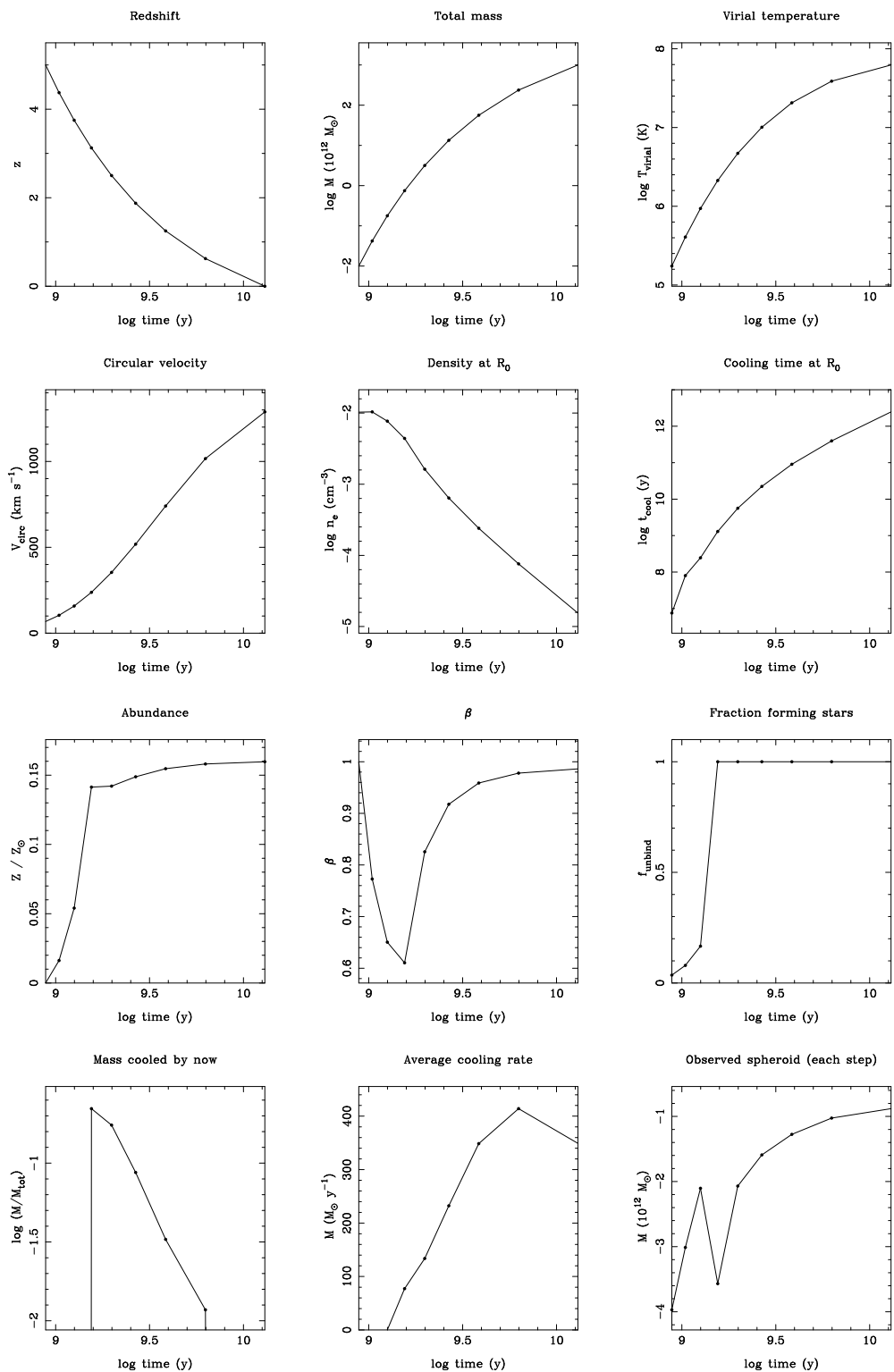

**Figure 1.** Collapse properties as a function of time for the default model. All quantities are plotted as a function of the log of the time in years. From left to right and top to bottom (a) shows: redshift; total mass of the system; virial temperature; circular velocity; electron number density of the gas at the truncation radius, $R_0$; cooling time of the gas at $R_0$. Panel (b) shows: the heavy element (iron) abundance at the start of each collapse step; $\beta$ for the hot gas; $f_{\rm unbind}$, the fraction of the catastrophically cooled gas which forms into stars during each collapse step; the mass of gas which would cool by the present (as a fraction of the total mass) if no further collapse occurs after each step; the average cooling rate in any cooling flow at each step; the mass of gas which turns into normal stars at each step. On the left in panel (c) we show the division of the baryonic mass, expressed as a fraction of the total mass, into gas, dark matter and normal stars at the end of each collapse step. The lowest curve gives the gas fraction. The difference between the lowest and middle curves is the fraction of baryonic dark matter (gas which has cooled to low temperature in a cooling flow). The difference between the middle and top curves is the fraction of matter in normal stars. In the right panel we show the spatial structure of the collapsed system at each step. The top curve shows the truncation radius, $R_0$. The bottom curve shows the outer boundary of the region where the gas cools catastrophically, $R_{\rm CF}$, forming the spheroid. The middle curve shows the outer boundary of the gas which cools during each collapse step.

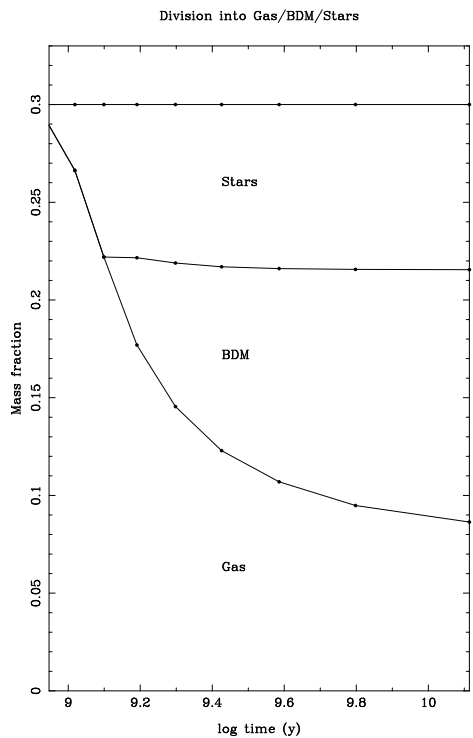
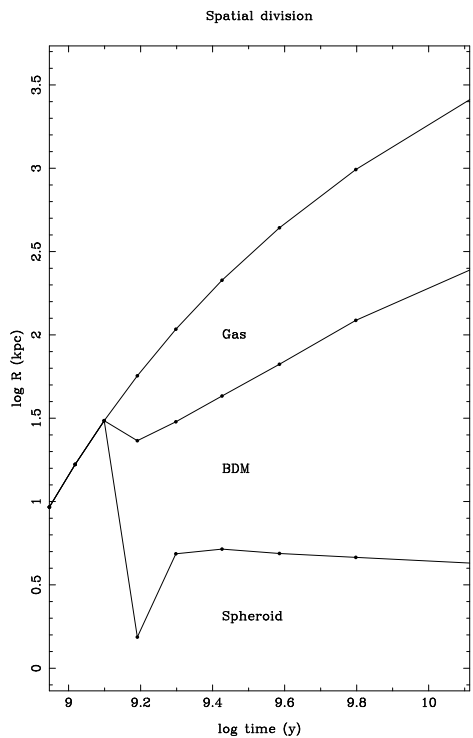

It has been argued elsewhere (Fabian & Nulsen 1994; see also Nulsen, Stewart & Fabian 1984) that the effective viscosity of the hot gas is always likely to be appreciable. Thus the hot gas tends to corotate, and as cooling gas flows inward it leaves it angular momentum behind in the remaining gas. Initially the specific angular momentum of the gas is likely to be negligible (Efstathiou & Jones 1979) but when most of the gas has cooled, leaving the bulk of the angular momentum in a small quantity of gas, the cooling flow will be significantly aspherical. Thus, the last of the cooling gas settles to form a disk. The mass of the disk is determined primarily by the angular momentum in the hot gas when the system collapses.

Note that the supernovae will blow away a cooling flow if that is possible. In our notation, this occurs when $f_{\rm unbind} \leq 1$. Since $f_{\rm unbind}$ is also the fraction of the prespheroid gas which gets turned into stars, we make it saturate at 1. Thus, in terms of our figures, a disk can only form in those systems for which $f_{\rm unbind} = 1$.

Thus, in order to form a disk a protogalaxy-galaxy must be large enough to form a cooling flow, but small enough for all the gas to have cooled by the present. In the default model of Fig. 2, for the stage that collapses in a halo of $1.6 \times 10^{12}\,{\rm M}_\odot$ the cooling time at the outer edge of the gas is $2.5 \times 10^9$ y, allowing the disk to have formed by about $4.2 \times 10^9$ y, or a redshift of about 1.

In this model, small systems which contain no cooling flow cannot form disks. Systems which merge with another system of comparable size or larger before the cooling finishes get stripped of their remaining gas and so do not have disks. However, disks can form around a large field elliptical galaxy (a large spheroid), but only where cooling has gone to completion. The presence of hot gas around many such objects (Forman, Jones & Tucker 1985) tells us that this has yet to happen.

### 3.3 Dwarf galaxies

Since the gas which does not form stars gets driven from small systems, they should all be spheroidal according to our model. In practice some gas may be left after the bulk of it has been expelled. The model is not much affected by this, provided that most of the gas is expelled and that star formation is greatly inhibited by the expulsion. The model says nothing about what determines the type of a dwarf galaxy.

There is no mechanism for disk formation in dwarf galaxies in this model. The lack of cooling flows in dwarf systems also means that they will not have baryonic dark haloes (see below).

### 3.4 Clusters of galaxies

In the default model, if the hierarchy proceeds to the mass of a cluster ($M \sim 10^{15}\,{\rm M}_\odot$) by about the present, the gas fraction is about 9 per cent and the iron abundance 0.16. Both of these values is too small to agree with measurements for present-day clusters. The metallicity can be increased by demanding that the yield per supernova is higher. This does not seem unreasonable considering the range of iron yield as a function of supernova mass shown in Renzini et al (1993).

It is apparent from the figures that most of the gas is consumed at about the stage when "normal" galaxies form. Lower mass systems are prevented from consuming much gas by supernova feedback. Higher mass systems have long cooling times (even ignoring the reduced gas fraction). A major shortcoming of our simple model is the assumption that each stage of the collapse hierarchy is formed by coalescing objects from the preceding stage. As a result, no object larger than a "normal" galaxy can have a gas fraction of more than about 18 percent. In reality most objects form by the merger of subsystems with a wide range of sizes. Many of these systems may be more gas rich than those which have formed a normal galaxy. Thus we would expect a more realistic merger model to produce clusters with higher gas fractions (also reducing the metallicity to some extent).

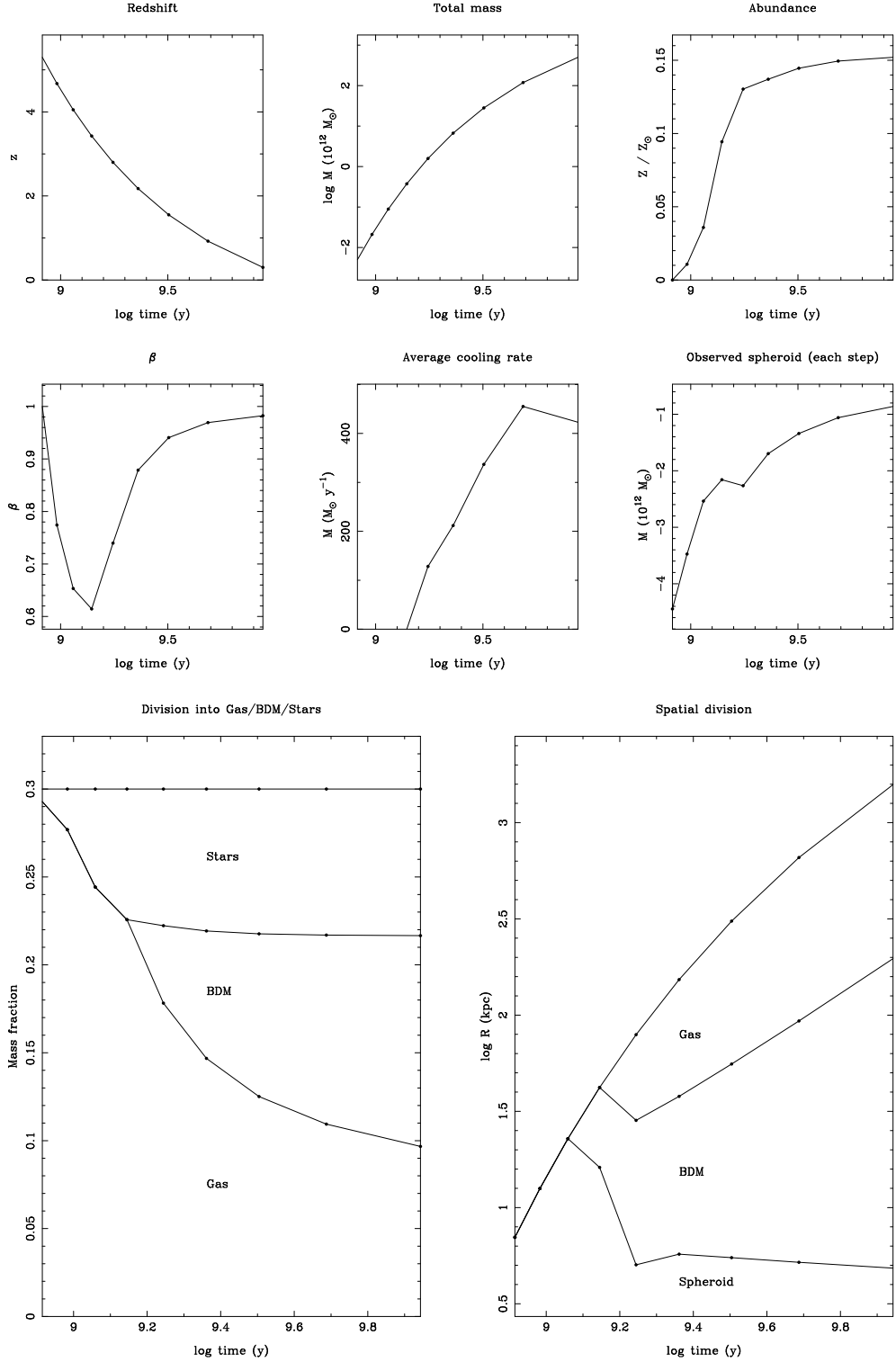

**Figure 2.** Sensitivity of the spheroid in "normal" galaxies to collapse details. Various properties are shown as functions of the time for a model which parallels the growth of the default model, with the collapses occurring at different times (and masses). All quantities are plotted as a function of the log of the time in years. From left to right and top to bottom (a) shows: redshift; total mass of the system; iron abundance at the start of each collapse step; $\beta$ for the hot gas; the average cooling rate in any cooling flow at each step; the mass of gas which turns into normal stars at each step. Panel (b) shows the same quantities as Fig. 1, panel (c).

### 3.5 Other parameter sets

We have carried out several runs varying parameters from the default set. The cooling flow phase starts earlier (and thus at lower mass) if: a) $\tau_0$ is reduced (thus $R_{\rm CF}$ is smaller and $f_{\rm unbind}$ reaches unity earlier); b) $f_{\rm B}$ is smaller (so the densities are lower and cooling times longer); or c) $z_{\rm start}$ is smaller. Forcing $\beta$ to be unity for all stages, as assumed for most previous work, has the effect of delaying the formation of a cooling flow phase. Fig. 3 shows the results of a simulation with $\beta = 1$. Note the massive spheroid produced in this run.

We note that our assumed cosmological model, with $\Omega_0 = 1$ and $f_{\rm B} \sim 0.3$ initially, is inconsistent with current models of cosmic nucleosynthesis (Walker et al. 1991). This is the problem of baryon overdensity in clusters (see S. White et al 1993; D. White & Fabian 1995) for which no convincing solution has yet been found. Reducing $\Omega_0$ forces consistency with the nucleosynthesis results but creates other problems for structure formation, which must then finish by a redshift of a few, contrary to observations of structure and evolution of clusters (see discussion by Richstone, Loeb & Turner 1992). A more complex cosmological model (say with a cosmological constant) will not eliminate the cooling flow phase but may change the mass scale for which it occurs.

## 4 DISCUSSION AND CONCLUSIONS

The present work shows that cooling flows play an important role in the formation of galaxies of mass comparable to the Galaxy and greater. Energy injected from supernovae associated with the massive stars formed in the catastrophically cooled gas, from where $\tau < \tau_0$, makes gas at the next stage of the hierarchy less dense than it would otherwise be, so promoting a cooling flow phase (where $\tau > \tau_0$) at lower galaxy masses. Indeed, in our model the formation of a cooling flow in much of the gas and the consequent quenching of massive star formation is responsible for the turnover in the galaxy luminosity function (*i.e.* at $L \sim L^*$). If normal star formation continued in the cooling flow gas, then there would be many more luminous galaxies visible today, unless the gas fraction is much lower than we have assumed (contrary to the observed content of clusters) or that scale is somehow fixed by the initial conditions.

A cooling flow of the magnitude produced by our default model during the formation of the Galaxy would have formed a baryonic dark halo containing about 20 percent of the total mass. This is roughly consistent with results from the search for MACHOS. These results indicate that about 20 percent of the dark halo of our Galaxy may be in the form of low-mass stars (Alcock et al 1993; Auborg et al 1993; Alcock et al 1995), probably brown dwarfs. Cluster cooling flows deposit cooled gas with a profile $M(<R) \propto R$ which, if followed by low-mass stars formed from a cooling flow in the halo, also follows the isothermal density distribution of the halo. The fact that observed limits on faint red stars in the galactic halo do not extrapolate to explain MACHOS (Bahcall et al. 1994) is due to the objects formed in cooling flows being an old population distinct from the bulge or disk populations of hydrogen burning stars. We see that for galaxies in general, a cooling flow phase leads to a smooth transition in the rotation curve from spheroid to dark halo, since the baryonic content originated from the same gas.

As discussed above, the disks of spiral galaxies form last, from the tail end of the cooling flow. It is well known that substantial dissipation is required to produce a galactic disk (Fall & Efstathiou 1980). Given the minimum collapse factor of 10 required for this (by the conventional mechanism) the material which formed the disk of the Galaxy must have fallen in from about 100 kpc or more. The first of the collisions required for dissipation must therefore have occurred while this material was very tenuous. It seems likely that such tenuous gas would have been slow to cool after collisions, so that it would have gone through a prolonged high temperature phase. In short, it is hard to avoid a high temperature phase during the formation of a (large) galactic disk. We argue that disk formation will always involve a cooling flow phase. Late formation of the disk clearly has some bearing on the G dwarf problem since the metallicity always exceeds 0.05 during the cooling flow phase.

There are two effects which cause $f_{\rm unbind}$ to increase through the collapse hierarchy: increasing binding energy of the gas and decreasing size of $R_{\rm CF}$. In our default model it is primarily the extra weight of the cooling flow gas (*i.e.* decreasing $R_{\rm CF}$) which makes $f_{\rm unbind}$ exceed 1. If instead $R_{\rm CF}$ remains large, then a larger spheroid will be formed in a deeper potential well at the stage that $f_{\rm unbind}$ reaches 1. (By our assumptions $f_{\rm unbind}$ will always be smaller than 1 until a cooling flow forms, but in a deeper potential a small cooling flow has a greater effect on $f_{\rm unbind}$.) An object where this occurred would resemble an elliptical galaxy. In order for it to form in our model the collapse needs to occur earlier. Fig. 4 shows a run which produces an elliptical galaxy at about $7 \times 10^8$ y (z = 5.8) in a halo with a total mass of $6.5 \times 10^{11}$ M$_\odot$. The iron abundance in this elliptical is 0.17, somewhat larger than at the end of our default run. It is possible that ellipticals are most common in clusters just because they formed from perturbations which collapsed early due to being part of a larger-scale cluster perturbation.

The elliptical galaxy model of Fig. 4 has one shortcoming, the short cooling time at the outer edge of the gas (about $4 \times 10^8$ y). Based on the arguments above, we should expect a disk to form in this system. Perhaps the supernova heating is sufficient to prevent cooling before now (since $2 \times 10^{10}$ M$_\odot$ of normal stars form in this spheroid).

We showed in a previous paper (Fabian & Nulsen 1994) that thermal conduction does not suppress the thermal instability necessary for widespread mass deposition by a cooling flow in systems of the size of the Galaxy. Along with heavy elements, early supernovae also inject magnetic fields into the gas, further reducing thermal conduction. Thus, as far a thermal stability is concerned, the situation in the ear-liest cooling flows is probably similar to that in rich clusters today.

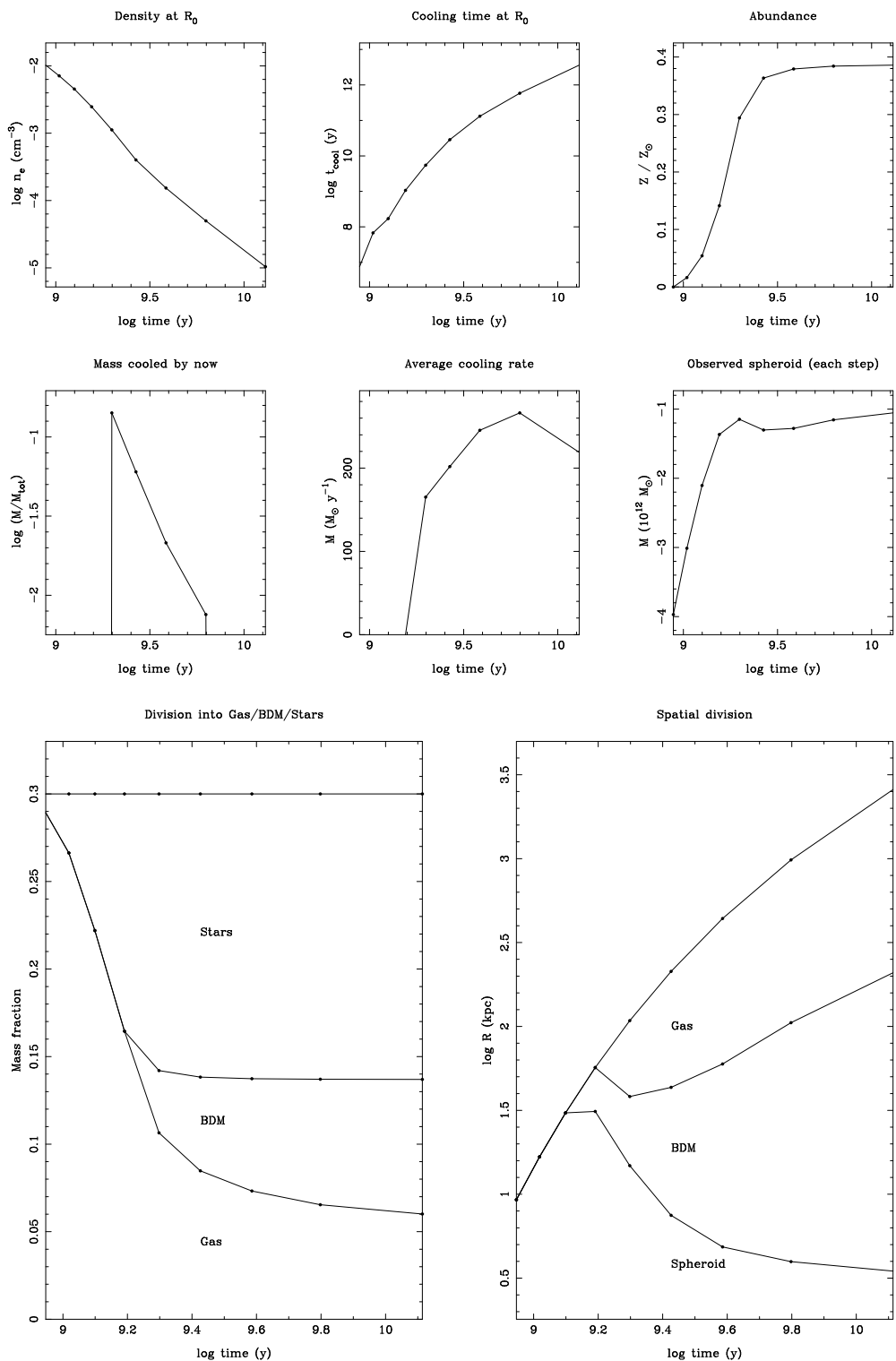

**Figure 3.** Effect of forcing $\beta = 1$. This shows various properties as a function of the time for a collapse model which is identical to the default model, except that $\beta$ for the hot gas was forced to be 1 at every step. From left to right and top to bottom panel (a) shows: electron number density of the gas at the truncation radius, $R_0$; cooling time of the gas at $R_0$; iron abundance at the start of each collapse step; the mass of gas which would cool by the present (as a fraction of the total mass) if no further collapse occurs after each step; the average cooling rate in any cooling flow at each step; the mass of gas which turns into normal stars at each step. Panel (b) shows the same quantities as Fig. 1, panel (c).

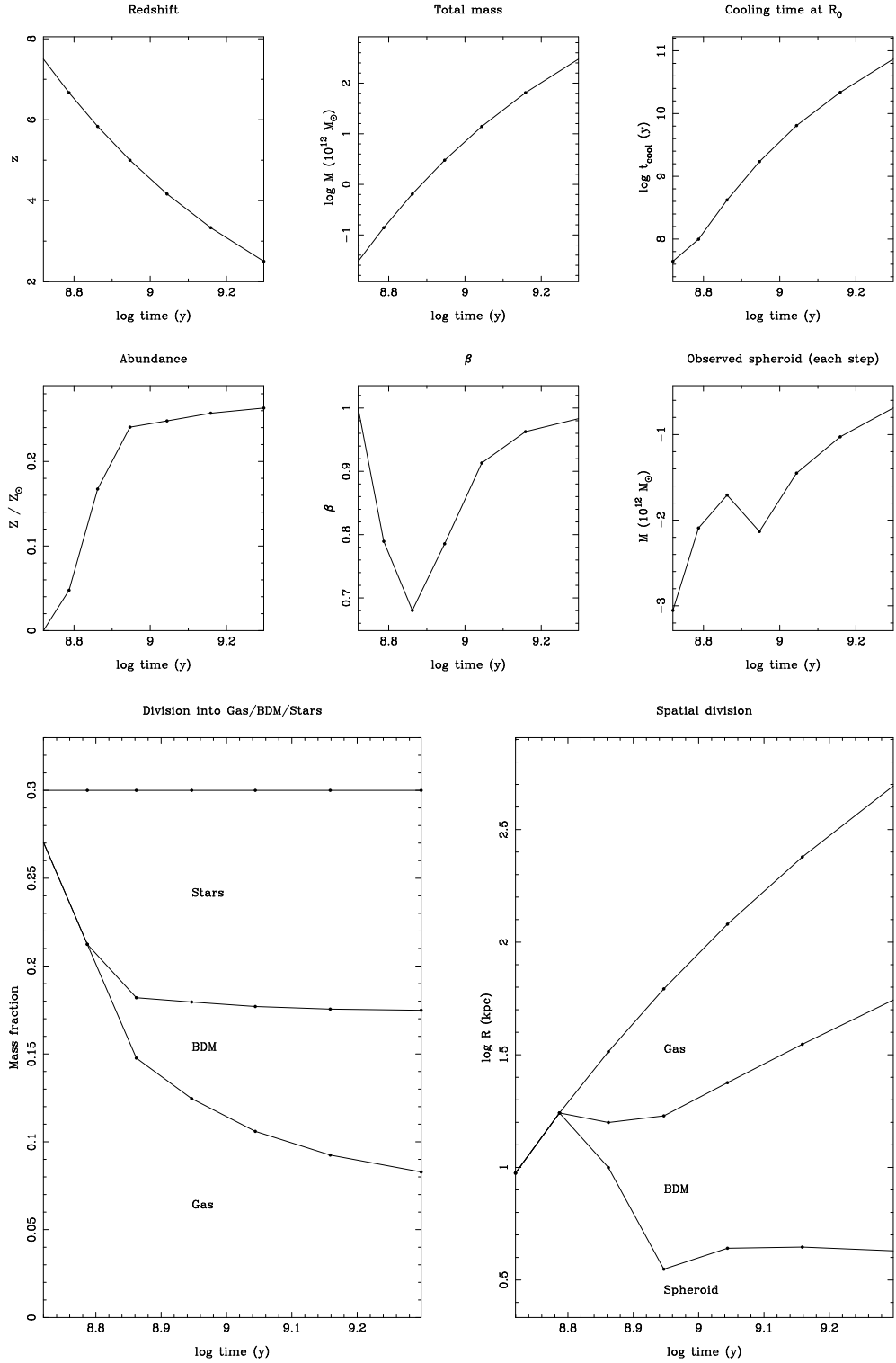

**Figure 4.** Early collapse produces a large spheroid. This collapse has the same parameters as the default model, but occurs earlier. From left to right and top to bottom panel (a) shows: redshift; total mass of the system; cooling time of the gas at the truncation radius; iron abundance at the start of each collapse step; $\beta$ for the hot gas; the mass of gas which turns into normal stars at each step. Panel (b) shows the same quantities as Fig. 1, panel (c).

One recurring problem in other simulations of galaxy formation (*e.g.* White & Frenk 1991; Cole et al 1994) is the tendency to overproduce low mass galaxies. Many small haloes collapse early during galaxy formation and each contains a small spheroid (in our terminology). Previous workers have assumed that these spheroids are bound sufficiently tightly that they do not "dissolve" along with their parent haloes at the next stage of the merger hierarchy. In that case, merging of the spheroids has to rely on dynamical friction which is very slow for small masses (*e.g.* see White & Frenk 1991). As noted above, the spheroids in our model do not have the opportunity to become very tightly bound, so that they are more prone to disruption than has been assumed.

The feature that distinguishes "normal" galaxies from the early spheroids is the cooling flow phase. It is this phase which offers the best opportunity for substantial dissipation by the gas (in principle the radiative heat loss from a self-gravitating cloud is almost arbitrary). With about 20 percent of the mass starting as gas, the binding energy released during the deposition of the baryonic dark matter could well be the key to the ability of galaxies to survive subsequent mergers. This phase is missing for the smaller systems, and makes a good candidate for conferring special survival status on "normal" galaxies.

The mass deposition rates in the cooling flows leading to the baryonic haloes of galaxies such as our own were about 100 $M_\odot$ yr$^{-1}$ at $z \sim 2-4$. This is similar to the rates of mass deposition around the central galaxies in many present-day clusters. The accumulation of small cold gas clouds in such flows may be associated with the observed damped Lyman-$\alpha$ clouds at those redshifts; the cooling gas in the residual flows at lower redshifts may be associated with the observed metal-line clouds, which require large haloes to galaxies.

One outcome which we might have anticipated is that supernova heating of the gas could account for the "beta problem" in clusters (Sarazin 1988), *i.e.* the excess energy per unit mass in the intergalactic medium. If anything, we have overestimated the effectiveness of supernova heating, and yet beta always ends up very close to 1 for cluster sized systems. It appears that supernova heating has a negligible effect on the thermal energy of the intracluster gas.

More detailed and, perhaps, more realistic cosmological models, using stochastic merger histories for galaxies and clusters will be explored in later papers. We first need a model which is consistent with all observations: the baryon overdensity in clusters; their substructure and recent evolution; and cosmic nucleosynthesis.

## 5 ACKNOWLEDGEMENTS

We thank Professor D. Wickramsinghe and the ANUATC for hospitality and support over the period during which much of the work was carried out. ACF also thanks the Royal Society for support.